# Single-Photon Molecular Cooling

Edvardas Narevicius[1,2]*, S. Travis Bannerman[2] and Mark G. Raizen[2]

[1]Department of Chemical Physics, Weizmann Institute of Science,
Rehovot 76100, Israel

[2]Center for Nonlinear Dynamics and Department of Physics,
The University of Texas at Austin, Austin, Texas 78712, USA

* edn@weizmann.ac.il

## Abstract

We propose a general method to cool the translational motion of molecules. Our method is an extension of single photon atomic cooling which was successfully implemented in our laboratory. Requiring a single event of absorption followed by a spontaneous emission, this method circumvents the need for a cycling transition and can be applied to any paramagnetic or polar molecule. In our approach, trapped molecules would be captured near their classical turning points in an optical dipole or RF-trap following an irreversible transition process.

Cooling the translational motion of molecules has been one of the grand challenges of physics and chemistry for many years, motivated by the wide range of fundamental problems that could be addressed. For example, the study of chemical reactions at ultra low temperatures would open a new and unexplored regime of cold chemistry, where reactions might proceed through resonance states [1] and the reaction rate and outcome could be controlled by externally applied fields [2]. Cold molecules would enable precision molecular spectroscopy, possibly allowing the observation of a variation in time of the fundamental constants [3-5]. Ultimately, cooling of molecules could enable the study of quantum degenerate polar molecular gases [6,7].

The established methods of molecular cooling work down to a temperature of tens to hundreds of mK. These include cooling by collisions with a cold buffer gas either in a cryogenic cell or during an adiabatic (supersonic) expansion. Buffer gas cooled paramagnetic molecules have been magnetically trapped [8] whereas cold supersonic molecular beams have to be first decelerated to trapping velocities by interaction with time dependent electric [9-11] or magnetic fields [12-16]. To reach lower temperatures and higher phase space densities, a different method must be used. A natural candidate is evaporative cooling, however this method is not likely to work for molecules in a low-field seeking state due to inelastic collisions. Application of standard laser cooling to molecules has been investigated by Di Rosa [17], however the method is not general due to the complicated internal structure. More general cooling methods have been suggested, including cavity cooling [18,19] and the ``optical shaker'' [20].

A new approach, single-photon cooling, was recently proposed and demonstrated by us for magnetically trapped atoms [21,22]. We show in this Letter that this method can be naturally extended to molecules. We start with an example of a straightforward application of the single photon cooling technique on a molecular radical and propose a new scheme that overcomes some of the deficiencies of the optical trap based single photon cooling method.

Optical tweezer based single photon molecular cooling begins with a far detuned optical dipole trap placed at the location of the classical turning point of the "hottest" molecules in the trap. As the "hottest" molecules arrive at the classical turning point, they exchange their kinetic energy for potential energy and come to near rest. At that point, we induce a resonant transition to a single rovibrational level of the electronically excited molecule. It decays back to the rovibrational ground state, with a certain probability given by the branching ratios (defined by the Hönl-London and Franck-Condon factors). The selection rules for the total angular momentum projection change in a spontaneous decay are $\Delta m_J = 0, \pm 1$. In the case where $\Delta m_J = -1$ the slope due to the magnetic quadrupole becomes smaller in the final state (assuming that the Lande factor does not change) and the potential well due to the attractive standing wave is deep enough to support a bound state. Since the lifetimes of electric-dipole allowed transitions range from tens to hundreds of nanoseconds, both the position and the kinetic energy of the molecule do not change during the decay (up to a photon recoil). If the spontaneous emission occurs at the location of the optical tweezer, the stationary molecule is trapped. This condition can be fulfilled by making the excitation volume small compared to the optical trap dimensions. The spontaneous decay into the ground state is irreversible since the final state is detuned from the initial level by the rotational energy level separation. In order to accumulate more molecules, we translate the center of the magnetic trap relative to the optical trap and the resonance excitation beam, thus picking up more molecules along the sweep.

We now analyze a single photon cooling processes of the NH radical, that has a $^3\Sigma^-$ electronic configuration in the ground state. The level diagram is displayed in Fig. 1a. Since the orbital momentum of an NH radical is zero in the $^3\Sigma^-$ state it has a particularly simple Zeeman level structure. The ground state NH molecules can be created in an arc discharge that would dissociate a precursor molecule such as $NH_3$ near the orifice of a supersonic nozzle. Magnetically trapped ground level molecules can be optically pumped into the rotationally excited initial level $|v=0, J=1, N=2\rangle$ (v, J and N are the vibration, total rotation and nuclear rotation quantum numbers respectively). We estimate that the optical pumping efficiency via the $|v=0, J=1\rangle$ level of the $^3\Pi_0$ manifold can be as high as 64%, using the absorption and fluorescence branching ratios derived from the Hönl-London factors for $^3\Pi \leftarrow {}^3\Sigma^-$ transition given in reference [23]. The projection of the total angular momentum on the quantization axis in the initial state is $m_{Ji}=1$. In order to confine molecules with a temperature of 50 mK to a 5 mm diameter cloud we need to apply a magnetic quadrupole field with a gradient of 0.14 T/cm. Since the polarizability of an NH radical is only $1.3*10^{-24}$ cm$^3$ we need to apply a high intensity optical field to create a deep enough attractive potential. A far detuned standing wave with 1kW of optical power focused to 100 μm beam waist will form a 250 μK deep well. The high optical intensity needed can be achieved by using a buildup cavity with a moderate finesse. We illustrate the combined magnetic and optical potential for the initial level in Fig 1b. The magnetic potential is so steep that the contribution from the optical potential is hardly noticeable. The irreversible transition from the $|v=0, J=1, N=2\rangle$ level is initiated by electronic excitation along the $^3\Pi \leftarrow {}^3\Sigma^-$ transition at 336 nm. We selectively populate the $|v=0, J=1\rangle$ level of the $^3\Pi_1$ manifold that decays into the ground $|v=0, J=1\ N=0, m_J=0\rangle$ sublevel with a probability of ~35%. Since the projection of the total angular momentum is zero in the final ground level, the magnetic contribution to the overall potential becomes negligible (see Fig 1c) and the NH radical becomes trapped in the standing optical wave. By continuing the cooling process, ground state NH radicals will accumulate in a shallow optical trap. Single-photon cooled and optically trapped molecules can be further cooled by evaporation. Evaporative cooling can be initiated by reducing the power of the standing wave. Since the molecules are in the overall ground state there will be no losses associated with inelastic collisions. Note, that technique was proposed to allow for multiple loadings of NH radicals into a magnetic trap [24].

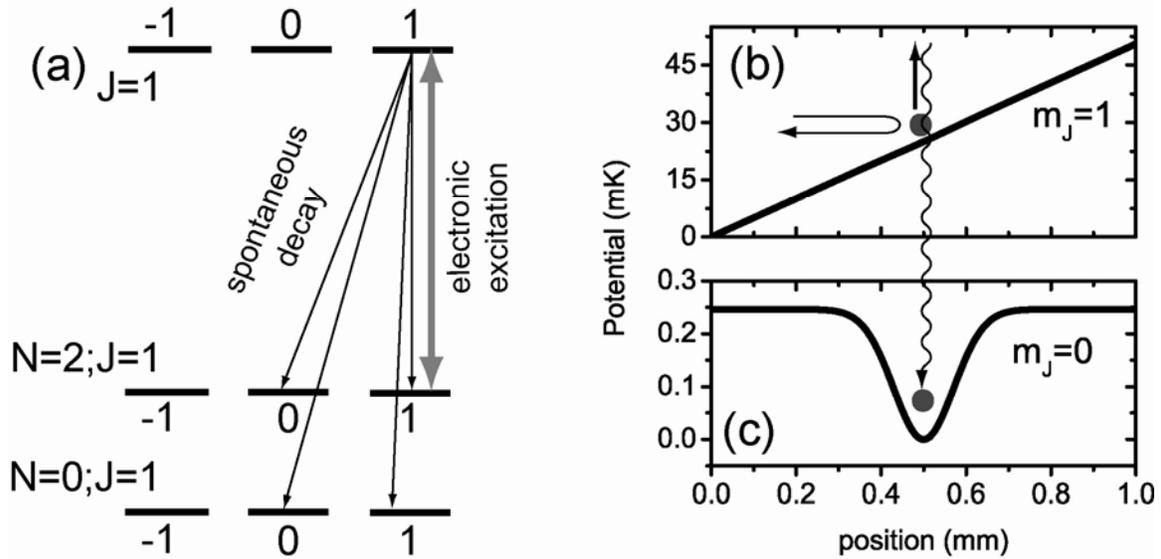

**Figure 1.** (a) The level diagram involved in the single photon cooling of NH radical. The combined magnetic and optical potentials in the initial (b) and final (c) levels of the cooling process. In the final level the projection of the total angular momentum is zero and only the optical tweezer contributes to the total potential.

During the single photon cooling process cold molecules accumulate in the tweezer after the irreversible transition that lowers the projection of the total angular momentum. The tweezer loading rate is limited by the frequency that molecules encounter the tightly focused electronic excitation beam. It is possible to enhance the loading rate using a different scheme based on accumulation of cold molecules in a trap formed by coupling different magnetic sublevels with an rf-field

A shell-like potential trap for neutral atoms formed by an avoided crossing in the RF-dressed magnetic sublevels was proposed by Zobay and Garraway [25]. Colombe et. al [26] have demonstrated atom loading into the rf-trap from the magnetic quadrupole trap and Schumm et. al. [27] demonstrated the formation of a double well RF-trap potential on an atom chip. We will show that single-photon cooling can be used to cool molecules by transfering the cold molecules into a local RF-trap. We use an example of a molecule in the $^2\Sigma$ state, particularly a CN radical.

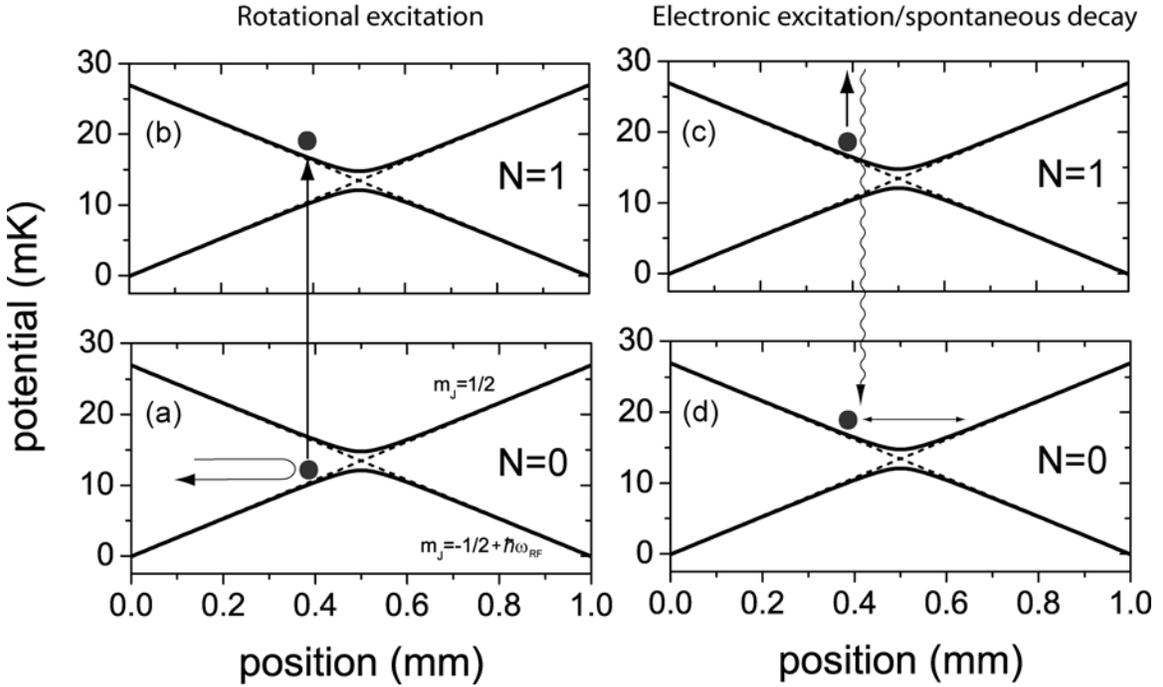

**Figure 2.** The schematic picture of the RF-trap based cooling process. (a) molecule reaches a classical turning point on the lower adiabatic potential surface. (b) the molecule is rotationally excited at the position of the classical turning point. (c) electronic excitation by a two photon process. (d) the molecule undergoes a spontaneous decay and becomes trapped on the upper adiabatic potential surface (RF-trap) of the rotationally ground N=0 manifold.

We consider a $^2\Sigma$ molecule trapped in a magnetic quadrupole trap. The ground rovibrational level, J=1/2, under the influence of magnetic field splits into two sublevels with a total angular momentum projection of $m_J$=1/2 and $m_J$=-1/2. Two magnetic sublevels can be coupled by the RF field, where the crossing point of two dressed potentials can be varied by adjusting the RF field frequency. At the crossing point the degeneracy is removed by the RF-coupling and two adiabatic potential surfaces with an avoided crossing are formed. As shown in Fig 2a the upper potential surface has a potential minimum at the crossing point. However, the turning point of a ``hot'' molecule is located on the lower adiabatic potential surface (see Fig 2a). A stationary molecule at the turning point can be transferred to the upper adiabatic potential surface by a two step irreversible transition. In the first step we rotationally excite a molecule at the classical turning point by coupling the initial, $N=0$, $J=1/2$, $m_J=1/2$, and final, $N=1$, $J=3/2$, $m_J=-1/2$, levels with a circularly polarized microwave field. The transition frequency can be tuned around the rotational transition at 113.9 GHz to selectively excite molecules at the classical turning point located in the vicinity of an RF avoided crossing. In a second stage we induce a two photon ($\lambda=776$ nm) electronic excitation along the $X^2\Sigma^+ \leftarrow B^2\Sigma^+$ transition around 388 nm and couple the $J=3/2$, $m_J=-1/2$ level to the $J=3/2$, $m_J=-1/2$ level of the $B^2\Sigma^+$ state. The excited molecule then decays (in ~60 ns) to the upper adiabatic surface of the rotationally ground, $N=0$, state manifold completing the cooling cycle. (Note, that the molecule returns to the same parity state as in the initial level due to the two photon transition used.) Now the molecule is trapped in the RF-trap and is far detuned from both the microwave and double photon optical field frequencies. Importantly, the whole magnetic trap volume can be illuminated with the 776 nm light as only the molecules that were rotationally excited near the classical turning points on the iso-B surface will absorb the two photons. Instead of accumulating molecules into a small volume optical tweezer, we load the cold molecules into an RF-trap extending on a surface. The molecules have to be slow enough to follow the adiabatic potential curve, otherwise they might be lost due to a non-adiabatic transition. Larger splitting at the avoided crossing point allows higher kinetic energy molecules to stay in the RF-trap.

In order to estimate the amplitude of the magnetic component of the RF field needed we have to compare the coupling strength to the frequency associated with the passage time through the avoided crossing region. We estimate that for a CN radical moving at 1 m/s (corresponding to a temperature of 3 mK) and assuming that the magnetic field gradient in the quadrupole trap is 1000 G/cm, and the amplitude of the magnetic component of the RF field is 1 G, the coupling strength is larger by more than two orders of magnitude compared to the level splitting. The RF field needed can be created by an oscillating electric field with an amplitude of 100 V across two electrodes spaced by 6 mm. In order to accumulate molecules with different energies we have to sweep both the microwave and RF frequencies and move the location of the RF-induced trap to lower energies. The RF-trap "implodes" catching atoms or molecules near their classical turning points.

The loading rate into the optical tweezer can be estimated as $\rho$ $V_{ex}$ $\omega_{trap}$, where $\rho$ is the trapped molecule density, $V_{ex}$ electronic excitation region volume and $\omega_{trap}$ is the magnetic (or electrostatic) trap frequency. Clearly it is limited by the small excitation volume; assuming the excitation region dimensions given by 10 μm x 10 μm x 100 μm (excitation beam waist of 10 μm) and density of the trapped molecules to be $10^8$ cm$^{-3}$ the loading rate approaches a magnetic trap frequency that can be as high as few kHz. The RF-trap loading method overcomes this problem by exciting the molecules from the iso-B surface. This method should allow us to capture *nearly all the molecules* from the magnetic trap, limited only by the branching ratio of the irreversible step. Even if that branching ratio is not large, single-photon cooling can work well because a transition only needs to happen once. Finally, we note that the RF cooling method presented above can be applied to atoms as well.

We thank A. Libson for careful reading of the manuscript. MGR acknowledges support from the R. A. Welch Foundation and the Army Research Office.

[1] N. Balakrishnan and A. Dalgarno, Chem. Phys. Lett. **341**, 652 (2001).
[2] R.V. Krems, Phys. Rev. Lett. **93**, 013201 (2004).
[3] V.V. Flambaum and M.G. Kozlov, Phys. Rev. Lett. **99**, 150801 (2007).
[4] D. DeMille *et al.*, Phys. Rev. Lett. **100**, 043202 (2008).
[5] T. Zelevinsky, S. Kotochigova, and Jun Ye, Phys. Rev. Lett. **100**, 043201 (2008).
[6] J.M. Sage *et al.*, Phys. Rev. Lett. **94**, 203001 (2005).
[7] S. Ospelkaus *et al.*, Nature Physics, Published online: 22 June 2008.
[8] J.D. Weinstein *et al.*, Nature **395**, 148 (1998).
[9] H.L. Bethlem, G. Berden and G. Meijer, Phys. Rev. Lett. **83**, 1558 (1999).
[10] S.Y.T. van de Meerakker, *et al.*, Phys. Rev. Lett. **94**, 023004 (2005).
[11] B.C. Sawyer *et al.*, Phys. Rev. Lett. **98,** 253002 (2007).
[12] E. Narevicius *et al.*, Phys. Rev. A **77**, 051401(R) (2008).
[13] E. Narevicius *et al.*, New J. Phys. **9**, 358 (2007).
[14] E. Narevicius *et al.*, Phys. Rev. Lett. **100**, 093003 (2008).
[15] N. Vanhaecke *et al.*, Phys. Rev. A **75**, 031402(R) (2007).
[16] S.D. Hogan *et al.*, Phys. Rev. A **76**, 023412(R) (2007).
[17] M. D. Di Rosa, Eur. Phys. J. D **31**, 395 (2004).
[18] P. Horak *et al.*, Phys. Rev. Lett. **79,** 4974 (1997); V. Vuletić and S. Chu, Phys. Rev. Lett. **84**, 3787 (2000).
[19] B.L. Lev *et al.*, Phys. Rev. A **77**, 023402 (2008).
[20] I.Sh. Averbukh and Y. Prior, Phys. Rev. Lett. **94**, 153002 (2005).
[21] G.N. Price *et al.*, Laser Physics, **17**, 965 (2007).
[22] G.N. Price *et al.*, Phys. Rev. Lett. **100**, 093004 (2008).

[23] A. Schadee, Astron. Astrophys. **41**, 203 (1975); ibid. **41**, 213 (1975).

[24] S.Y.T. van de Meerakker *et al.*, Phys. Rev. A **64**, 041401(R) (2001).

[25] O. Zobay and B.M. Garraway, Phys. Rev. Lett. **86** 1195 (2001); Phys. Rev. A **69**, 023605 (2004).

[26] Y. Colombe *et al.*, Europhys. Lett. **67**, 593 (2004).

[27] T. Schumm *et al.*, Nature Physics **1**, 57 (2005).